\documentclass[twocolumn,pra]{revtex4}%
\usepackage{graphics,amsmath,amsfonts,amscd,revsymb,latexsym,
enumerate,multirow,epsfig}
\usepackage{amsmath}
\usepackage{amsfonts}
\usepackage{amssymb}
\usepackage{color}
\usepackage{graphicx}
\providecommand{\U}[1]{\protect\rule{.1in}{.1in}}
\providecommand{\U}[1]{\protect\rule{.1in}{.1in}}

\begin{document}
\preprint{ }
\title[ ]{Gapped spin Hamiltonian motivated by quantum teleportation}
\author{Ari Mizel}
\email{ari@arimizel.com}
\affiliation{Laboratory for Physical Sciences, 8050 Greenmead Drive, College Park,
Maryland, USA\ 20740}

\begin{abstract}
We construct a Hamiltonian whose ground state encodes a time-independent emulation of quantum teleportation.  We calculate properties of the Hamiltonian, using exact diagonalization and a mean-field theory, and argue that it has a gap.  The system exhibits an illuminating relationship to the well-known AKLT (Affleck, Lieb, Kennedy and Tasaki) model.  
\end{abstract}
\volumeyear{2014}
\volumenumber{ }
\issuenumber{ }
\eid{ }
\date{\today}




\maketitle

\section{Introduction}
Quantum teleportation \cite{Bennett1993} is an essential protocol in quantum information theory.  It is directly tied to a number of other protocols such as a ``superdense'' encoding scheme for data transmission \cite{Bennett1992},  a universal quantum computation method \cite{Gottesman99}, and a strategy for fault-tolerant quantum error correction \cite{Knill2005}.  It has also inspired productive subfields, such as studies of entanglement measures and studies of LOCC (local operations and classical communication) operations.

In this paper, motivated by quantum teleportation, we construct a parent spin Hamiltonian.  Its ground state is designed to provide a time-independent emulation \cite{Mizel02} of the standard time-dependent quantum teleportation protocol.  Formulated appropriately, the ground state possesses edge spins bound into an imperfect Bell pair.  We analyze the parent spin Hamiltonian using a mean field theory inspired by exact diagonalization results and obtain compelling evidence that the Hamiltonian is gapped.  Our system turns out to yield intriguing insights into the AKLT model \cite{Affleck1987,Affleck1988,Affleck1989}  and therefore into some general aspects of 1 dimensional antiferromagnetism \cite{Haldane1983}.

In the field of quantum information, spin chains have appeared in proposals for quantum buses and quantum channels \cite{Bose2003,Christandl2004,Verstraete2004,Wojcik2005,Venuti2006,Venuti2007,Cappellaro2007,Ferreira2008,Friesen2007,Wang2009,Wu2009,Oh2010,Yao2010,Banchi2011,Oh2011,Yang2011,Pemberton2011,Shim2011,Oh2012}.   It proves instructive to consider the relationship of our Hamiltonian  to some of these proposals as well.

Section II of the paper describes the Hamiltonian of the spin chain and its ground state.  The ground state includes a Bell pair with a member at each end of the chain, and we calculate the fidelity of this pair.  Section III of the paper computes the gap between the ground state of the chain and the excited states.  As expected from Lieb-Robinson arguments \cite{Hastings06}, the gap of the spin chain is inversely related to the fidelity of the Bell pair in the ground state.  We calculate this dependence quantitatively.  Section IV considers the relationship between our Hamiltonian and the AKLT model and also some of the quantum channel proposals.  Section V concludes.

\section{Hamiltonian}
\begin{figure}[htb]
\begin{center}
 \includegraphics[width=3.0in]{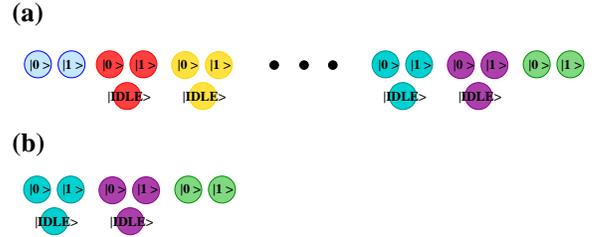}
 \end{center}
 \caption{(a) Spin chain formed by a line of qutrits (spin-1 particles) capped with qubits (spin-1/2 particles).  Chain has Hilbert space dimension $2 \otimes (3 \otimes 3)^{\otimes \ell}\otimes 2$; we refer to its length as $\ell$.   (b) Just the final unit of the chain, with Hilbert space dimension $3 \otimes 3 \otimes 2$.}
  \label{fig:spinchain}
\end{figure}

A diagram of a spin chain appears in Fig.~\ref{fig:spinchain}(a).  It has a Hilbert space of dimension $2 \otimes (3 \otimes 3) ^{\otimes \ell}\otimes 2 = 2 \otimes 3^{\otimes 2\ell}\otimes 2$.  We will refer to the length of the chain as $\ell$ since our construction presumes the $3$ dimensional spaces come in pairs; our construction does not permit a chain of Hilbert space dimension $2 \otimes 3^{\otimes 2\ell+1}\otimes 2$.  The chain can be regarded as a system composed of qutrits with a qubit capping each end or as a chain of spins of magnitude 1 with a spin of magnitude 1/2 capping each end.

The right-most unit, which has a $3 \otimes 3 \otimes 2$ dimensional Hilbert space, is depicted in Fig.~\ref{fig:spinchain}(b).  Its basis is defined by $\left\{ \left|0\right>, \left|1\right>,\left|{\scriptscriptstyle IDLE}\right> \right\} \otimes \left\{ \left|0\right>, \left|1\right>,\left|{\scriptscriptstyle IDLE}\right> \right\} \otimes \left\{ \left|0\right>, \left|1\right>\right\}$; the motivation behind the choice of ket label ${\scriptscriptstyle IDLE}$ will become clear.  We take the Hamiltonian to be $H(\theta) = I \otimes {\mathcal H}_{Create \,\, pair} + {\mathcal H}_{Projection}(\theta) \otimes I$ where
\begin{eqnarray}
\lefteqn{{\mathcal H}_{Create \,\, pair}  =} \nonumber\\
 &\epsilon& [(\left|1\right>\left|0\right>-\left|0\right>\left|1\right>)(\left<1\right|\left<0\right|-\left<0\right|\left<1\right|) \nonumber\\
&+&(\left|1\right>\left|0\right>+\left|0\right>\left|1\right>)(\left<1\right|\left<0\right|+\left<0\right|\left<1\right|) \nonumber\\
&+&(\left|0\right>\left|0\right>-\left|1\right>\left|1\right>)(\left<0\right|\left<0\right|-\left<1\right|\left<1\right|)]/2 \label{eq:HCreatepair}
\end{eqnarray}
and
\begin{eqnarray}
\lefteqn{{\mathcal H}_{Projection}(\theta) = } \nonumber\\
&\epsilon  & [\left(\sin \theta \frac{\left|0\right>\left|0\right>+\left|1\right>\left|1\right>}{\sqrt{2}} - \cos \theta \left|{\scriptscriptstyle IDLE}\right>\left|{\scriptscriptstyle IDLE}\right> \right) \nonumber \\
& & \;\;\;\left( \sin \theta \frac{\left<0\right|\left<0\right|+\left<1\right|\left<1\right|}{\sqrt{2}}- \cos \theta \left<{\scriptscriptstyle IDLE}\right|\left<{\scriptscriptstyle IDLE}\right|\right)\nonumber\\
& &+\sum_{b=0,1} \left|{\scriptscriptstyle IDLE}\right>\left<{\scriptscriptstyle IDLE}\right| \otimes  \left|b\right>\left<b\right| \nonumber \\
& & +\sum_{b=0,1}  \left|b\right>\left<b\right| \otimes \left|{\scriptscriptstyle IDLE}\right>\left<{\scriptscriptstyle IDLE}\right|] \label{eq:HProjection}
\end{eqnarray}
and where $\epsilon$ has units of energy.  One could imagine attempting to engineer this system using a line of quantum dots with a Hubbard Hamiltonian as in \cite{Shim2010}, but we have not carried out such an investigation.

\begin{figure}[htb]
\begin{center}
 \includegraphics[width=1.0in]{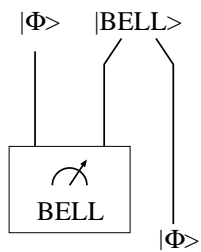}
\end{center}
 \caption{Quantum teleportation circuit, with time flowing downward.  To teleport an arbitrary state $\left| \Phi \right>$, one uses a Bell pair $(\left|0\right>\left|0\right>+\left|1\right>\left|1\right>)/\sqrt{2}$.  If a Bell-basis measurement of the arbitrary state and the left member of the pair yields the result $(\left|0\right>\left|0\right>+\left|1\right>\left|1\right>)/\sqrt{2}$, then the right member of the pair has state $\left| \Phi \right>$ after the measurement.}
  \label{fig:teleportcircuit}
\end{figure}

The Hamiltonian is motivated by the teleportation circuit \cite{Bennett1992,Bennett1993} shown in Fig.~\ref{fig:teleportcircuit}.  
${\mathcal H}_{Create \,\, pair}$ produces the Bell pair needed for teleportation while the projector ${\mathcal H}_{Projection}(\theta)$ performs a Hamiltonian analogue of the Bell-basis measurement.  To see this, note that the unit shown in Fig.~\ref{fig:spinchain}(b) has a doubly degenerate ground state
\begin{equation}
\left| \psi_0(b)\right>  = \frac{\begin{array}{c}\cos \theta \left|b\right>\otimes(\left|0\right>\otimes \left|0\right>+\left|1\right>\otimes \left|1\right>)/ \sqrt{2}  \\ 
+ \sin \theta \, \left|{\scriptscriptstyle IDLE}\right>\otimes\left|{\scriptscriptstyle IDLE}\right> \otimes \left|b\right>/2  \end{array}}{\sqrt{\cos^2\theta + (1/4) \sin^2\theta}} \label{eq:psi(b)}
\end{equation}
where $b$ can take either bit value $0$ or $1$.  The term proportional to $\cos \theta$ includes an initial state $ \left|b\right>$  alongside an initial Bell pair.  The form of this initial Bell pair is dictated by ${\mathcal H}_{Create \,\, pair}$.  The term proportional to $\sin \theta$ corresponds to teleportation of $ \left|b\right>$ after the Bell measurement.   This term arises because  ${\mathcal H}_{Projection}(\theta)$ uses an extra ``post-measurement'' state $\left|{\scriptscriptstyle IDLE}\right>\otimes \left|{\scriptscriptstyle IDLE}\right>$ to amplify the part of the $\cos \theta$ term that would result from a Bell-basis measurement with outcome $(\left|0\right>\left|0\right>+\left|1\right>\left|1\right>)/\sqrt{2}$ (i.e.  $(\left<0\right|\left<0\right|+\left<1\right|\left<1\right|)/\sqrt{2} \otimes I$ applied to the $\cos \theta$ term in (\ref{eq:psi(b)})).  Inspecting the form of (\ref{eq:psi(b)}), one sees that $\theta \approx 0$ produces negligible amplification while $\theta$ close to $\pi/2$ produces strong amplification and therefore successful teleportation.  Note that the 2 sums in ${\mathcal H}_{Projection}(\theta)$ impose an energy penalty unless two particles transition to the post-measurement state concomitantly.

For a chain of length $\ell$, the total Hamiltonian shown in  Fig.~\ref{fig:spinchain}(a) produces repeated teleportation down the chain by repeating $H(\theta)$.  The total Hamiltonian is
\begin{eqnarray}
{\mathcal H}(\theta) & = & {\mathcal H}_{Create \,\, pair} \otimes I^{\otimes 2\ell - 1}\otimes I \nonumber \\
& & + I\otimes \sum_{j=0,\ell-1} I^{\otimes 2j} \otimes H(\theta) \otimes I^{\otimes 2\ell - 2 - 2j}. \label{eq:Htotal}
\end{eqnarray}
Its ground state is most easily described by defining an operator $\hat{g}_0 = \left| \psi_0(0)\right>\left<0\right|+\left| \psi_0(1)\right>\left<1\right|$ in terms of (\ref{eq:psi(b)}).  This operator is a map from a $2$ dimensional Hilbert space to a $3 \otimes 3 \otimes 2$ dimensional Hilbert space.  In terms of $\hat{g}_0$, the ground state has the form
\begin{eqnarray}
\lefteqn{\left|\Psi\right> =  (I \otimes I^{\otimes 2 \ell -2} \otimes \hat{g}_0) \dots} \nonumber \\
& & (I \otimes I^{\otimes 2} \otimes \hat{g}_0) (I \otimes \hat{g}_0) \frac{\left|0\right>\left|0\right>+\left|1\right>\left|1\right>}{\sqrt{2}}. \label{eq:Psi}
\end{eqnarray}
To verify that the qubits capping the ends of the chain are entangled into a Bell-pair, we trace out the qutrits from the density matrix  $\mbox{Tr}_{3 \otimes 3}  \dots \mbox{Tr}_{3 \otimes 3} \left|\Psi\right>\left<\Psi\right|$.  To evaluate this trace, define the superoperator $g_0 \left(\rho\right) = \mbox{Tr}_{3 \otimes 3}  \hat{g}_0 \rho \hat{g}_0 ^{\dagger} = ( \sin^2  \theta \rho + 4 \cos^2  \theta \, \mbox{Tr} \rho\,  I /2 )/(4 \cos^2 \theta + \sin^2 \theta)$.  This is a depolarizing channel that approaches perfect transmission as $\theta$ approaches $\pi/2$.  We find that  $\mbox{Tr}_{3 \otimes 3}  \dots \mbox{Tr}_{3 \otimes 3} \left|\Psi\right>\left<\Psi\right| = \sum_{b,b^\prime = 0,1} \left|b\right>\left<b^\prime \right| \otimes g_0 (g_0( \dots g_0(\left|b\right>\left<b^\prime\right|)\dots))/2 = (\sin^2 \theta/(4 \cos^2 \theta + \sin^2 \theta))^{\ell} \sum_{b,b^\prime = 0,1} \left|b\right>\left<b^\prime \right| \otimes \left|b\right>\left<b^\prime \right|/2  + (1 - (\sin^2 \theta/(4 \cos^2 \theta + \sin^2 \theta))^{\ell}) I \otimes I /4 $.  Thus, the qubits capping the ends are entangled into a Bell pair of density matrix $\sum_{b,b^\prime = 0,1} \left|b\right>\left<b^\prime \right| \otimes \left|b\right>\left<b^\prime \right|/2 = (\left|0\right>\otimes \left|0\right>+\left|1\right>\otimes \left|1\right>) (\left<0\right|\otimes \left<0\right|+\left<1\right|\otimes \left<1\right|)/2$ with a fidelity that decreases like  $f(\theta)^{\ell}$ with the length of the chain, where
\begin{equation}
f(\theta) = \sin^2 \theta/(4 \cos^2 \theta + \sin^2 \theta). \label{eq:f}
\end{equation}
This is the same fidelity that would be obtained by forming a pair on adjacent qubits and then swapping the quantum information in one member of the pair $\ell$ times down a chain with fidelity $f(\theta)$ per swap.

\section{Energy gap}

\begin{figure}[htb]
\begin{center}
\includegraphics[width=3.0in]{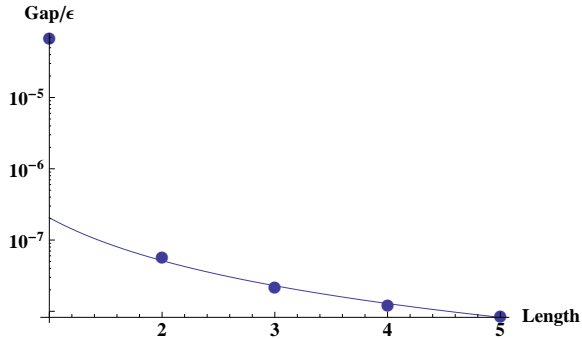}
\end{center}
\caption{Energy gap as function of chain length in units of the Hamiltonian energy scale $\epsilon$ appearing in (\ref{eq:HCreatepair}) and (\ref{eq:HProjection}).  Circles are the result of exact diagonalization, and solid line is a fit to $1/\ell^2$.  Pronounced length dependence is evident.}
\label{fig:gapversuschainlength}
\end{figure}

We now analyze the parent spin Hamiltonian (\ref{eq:Htotal}) and argue that it possesses a energy gap between its ground state and excited states (even as the length $\ell$ goes to infinity).  For $\theta = 0$, the Hamiltonian decouples into independent units, and it is evident from inspection that there is a gap of size $\epsilon$.  For $\theta$ approaching $\pi/2$, we first perform exact diagonalization of ${\mathcal H}(\theta)$ for small lengths, choosing $\theta = 1.56$ as a representative value.  (This leads to a fidelity value (\ref{eq:f}) greater than $0.9995$.)  A plot of the resulting gap, as a function of length $\ell$, appears in Fig.~\ref{fig:gapversuschainlength}.
This figure shows a pronounced dependence of the gap on $\ell$.  However, we do not believe that this dependence will persist once $\ell$ becomes very large; we attribute the dependence of gap of $\ell$ in Fig. \ref{fig:gapversuschainlength}  to finite-size effects that should die out for large $\ell$.  Unfortunately, it is infeasible to demonstrate this explicitly by performing exact diagonalization for substantially larger values of $\ell$ given the exponential growth in the Hilbert space as a function of $\ell$.  We therefore present evidence based on a mean field, variational framework.


To find an appropriate form for the variational state, we inspect our exact diagonalization results for $\ell=1$ to $\ell =5$.  For all these $\ell$, the first excited states comprise a degenerate triplet.  (The ground state of the chain is always found to take the form (\ref{eq:Psi}), as expected.)

We find that this excited triplet can be well described by states of the form
\begin{eqnarray}
\lefteqn{\left|\Psi_f\right>\propto  } \nonumber \\
& &\,\,\left[(I \otimes I^{\otimes 2 \ell -2} \otimes \hat{g}_f) \dots (I \otimes I^{\otimes 2} \otimes \hat{g}_0) (I \otimes \hat{g}_0)+ \dots \right. \nonumber \\
& &  + (I \otimes I^{\otimes 2 \ell -2} \otimes \hat{g}_0) \dots (I \otimes I^{\otimes 2} \otimes \hat{g}_f) (I \otimes \hat{g}_0) \nonumber \\
& &  \left. + (I \otimes I^{\otimes 2 \ell -2} \otimes \hat{g}_0) \dots (I \otimes I^{\otimes 2} \otimes \hat{g}_0) (I \otimes \hat{g}_f) \right] \nonumber\\
& &\hspace{2.0in}\frac{\left|0\right>\left|0\right>+\left|1\right>\left|1\right>}{\sqrt{2}} \label{eq:Psif}
\end{eqnarray}
where $\hat{g}_f = \left| \psi_f(0)\right>\left<0\right|+\left| \psi_f(1)\right>\left<1\right|$ is defined in analogy to $\hat{g}_0$.  For instance, in the case $\ell = 3$ and $\theta = 1.56$, we can choose three variational states $\left|\Psi_1\right>$, $\left|\Psi_2\right>$, and $\left|\Psi_3 \right>$ of the form (\ref{eq:Psif}) that overlap with the three exact triplet states to within $10^{-3}$ of unity.  To obtain this high overlap, for $\left|\Psi_1\right>$, we set $\hat{g}_f = \hat{g}_1  = \left|\psi_0(1)\right>\left<0\right| + \left|\psi_0(0)\right>\left<1\right|$; for $\left|\Psi_2\right>$, we set  $\hat{g}_f = \hat{g}_2  = \left|\psi_0(1)\right>\left<0\right| - \left|\psi_0(0)\right>\left<1\right|$; and, for $\left|\Psi_3\right>$, we set $\hat{g}_f = \hat{g}_3  = \left|\psi_0(0)\right>\left<0\right| - \left|\psi_0(1)\right>\left<1\right|$.  Equation (\ref{eq:Psif}) can be regarded as a ``spin-wave'' in which the excitation $\hat{g}_f$ is traveling down the chain.

Motivated by the spin-wave form (\ref{eq:Psif}), we choose the following ansatz for excited states of a long chain
\begin{eqnarray}
\lefteqn{\left|\Psi_{f,n}\right> = \frac{1}{\sqrt{\ell}} \sum_{j=0}^{\ell-1} \cos \frac{2\pi n j}{\ell} (I \otimes I^{\otimes 2 \ell -2} \otimes \hat{g}_0) \dots} \nonumber \\
& & (I \otimes I^{\otimes 2(j+1)} \otimes \hat{g}_0)(I \otimes I^{\otimes 2j} \otimes \hat{g}_f)(I \otimes I^{\otimes 2(j-1)} \otimes \hat{g}_0) \nonumber \\
& & \dots(I \otimes I^{\otimes 2} \otimes \hat{g}_0) (I \otimes \hat{g}_0) \frac{\left|0\right>\left|0\right>+\left|1\right>\left|1\right>}{\sqrt{2}}. \label{eq:Psifn}
\end{eqnarray}
The $\left| \psi_f(b)\right>$ appearing in $\hat{g}_f$ are to be determined by minimizing the expectation value of $\left<\Psi_{f,n}\right|{\mathcal H}(\theta)\left|\Psi_{f,n}\right> - E(\left<\Psi_{f,n}\middle|\Psi_{f,n}\right> -1) + \kappa(\sum_{b,b^\prime} \left|\left< \psi_f(b)\middle|\psi_0(b^\prime)\right>\right|^2)$.  The last term, which includes the Lagrange multiplier $\kappa$, is included as a means of imposing orthogonality between $\left| \psi_f(b)\right>$ and $\left| \psi_0(b^\prime)\right>$.  This orthogonality ensures that $\left|\Psi_{f,n}\right>$ is properly normalized by its $1/\sqrt{\ell}$ prefactor.  Indeed, when such orthogonality is not imposed, the resulting equations have solutions that are not properly normalized by this prefactor and exhibit strong end effects.   Eliminating end effects, which are present in the exact diagonalization calculations, is the goal of our variational framework, so all of the minimizations that we present in the following include the Lagrange multiplier term.

The minimization yields the variational equation
\begin{equation}
\frac{1}{2} \left[ \begin{array}{cc} H_{0,0}(\theta) & H_{0,1}(\theta) \\ H_{1,0}(\theta) & H_{1,1}(\theta) \end{array} \right]  \left[ \begin{array}{c} \left| \psi_f(0)\right> \\ \left| \psi_f(1)\right>\end{array} \right] = E_{f,n} \left[ \begin{array}{c} \left| \psi_f(0)\right> \\ \left| \psi_f(1)\right> \end{array} \right]. \label{eq:variational}
\end{equation}
In deriving this equation, we have defined
\begin{widetext}
\begin{eqnarray}
\lefteqn{H_{0,0}(\theta) = {\mathcal H}_{Projection}(\theta) \otimes I + \frac{4 \cos^2 \theta}{4 \cos^2 \theta + \sin^2 \theta} \left( I \otimes {\mathcal H}_{Create \,\, pair} + \frac{\left|0\right>\left<0\right|+ 2 \left|1\right>\left<1\right|}{2} \otimes I \otimes I \right.}   \nonumber \\
& &   + \cos \frac{2 \pi n}{\ell} (\left|0\right>\left<1 \right| \otimes \frac{\left| 0 \right> \left<0 \right| \otimes \left| 1 \right> \left<0 \right| +  \left|0\right>\left<1 \right| \otimes  \left| 1 \right> \left<1 \right|}{2}+\left|1\right>\left<0 \right| \otimes \frac{\left| 0 \right> \left<0 \right| \otimes \left| 0 \right> \left<1 \right| +  \left|1\right>\left<0 \right| \otimes  \left| 1 \right> \left<1 \right|}{2}  \nonumber \\
& & + \left|0\right>\left<0 \right| \otimes \frac{\left|0\right>\left<0 \right| \otimes \left|0\right>\left<0 \right| + \left|0\right>\left<1 \right| \otimes\left|0\right>\left<1 \right| - \left|1\right>\left<0 \right| \otimes \left|1\right>\left<0 \right| - \left|1\right>\left<1 \right| \otimes\left|1\right>\left<1 \right|}{4}  \nonumber \\
& & + \left. \left|0\right>\left<0 \right| \otimes \frac{\left|0\right>\left<0 \right| \otimes \left|0\right>\left<0 \right| + \left|1\right>\left<0 \right| \otimes\left|1\right>\left<0 \right| - \left|0\right>\left<1 \right| \otimes \left|0\right>\left<1 \right| - \left|1\right>\left<1 \right| \otimes\left|1\right>\left<1 \right|}{4} )\right) \nonumber \\
& & +\kappa ( \left|\psi_0(0)\right>\left<\psi_0(0)\right|+\left|\psi_0(1)\right>\left<\psi_0(1)\right|),
\end{eqnarray}
and
\begin{eqnarray}
\lefteqn{H_{0,1}(\theta) =\frac{4 \cos^2 \theta}{4 \cos^2 \theta + \sin^2 \theta} \left(-\frac{\left|0\right>\left<1\right|}{2}  \otimes I \otimes I \right.} \nonumber\\
& & + \cos \frac{2 \pi n}{\ell} (\left|0\right>\left<0 \right| \otimes \frac{\left| 1 \right> \left<0 \right| \otimes \left| 0 \right> \left<0 \right| +\left|1\right>\left<1 \right| \otimes \left| 0 \right> \left<1 \right|}{2}+\left|1\right>\left<1 \right| \otimes \frac{\left| 0 \right> \left<0 \right| \otimes \left| 0 \right> \left<1 \right| +\left|1\right>\left<0 \right| \otimes \left| 1 \right> \left<1 \right|}{2} \nonumber \\
& & - \left|0\right>\left<1\right| \otimes \frac{\left|0\right>\left<0 \right| \otimes \left|0\right>\left<0 \right| + \left|0\right>\left<1 \right| \otimes\left|0\right>\left<1 \right| - \left|1\right>\left<0 \right| \otimes \left|1\right>\left<0 \right| - \left|1\right>\left<1 \right| \otimes\left|1\right>\left<1 \right|}{4} \nonumber \\
& &  \left.- \left|0\right>\left<1\right| \otimes \frac{\left|0\right>\left<0 \right| \otimes \left|0\right>\left<0 \right| + \left|1\right>\left<0 \right| \otimes\left|1\right>\left<0 \right| - \left|0\right>\left<1 \right| \otimes \left|0\right>\left<1 \right| - \left|1\right>\left<1 \right| \otimes\left|1\right>\left<1 \right|}{4}) \right)
\end{eqnarray}
\end{widetext}
and have ignored small corrections at the ends of the chain that vanish in the limit $\ell \gg 1$.  We have also defined operators $H_{1,0}(\theta)$  and $H_{1,1}(\theta)$, which are given by taking $\left|1\right> \leftrightarrow \left|0\right>$ in $H_{0,1}(\theta)$ and $H_{0,0}(\theta)$ respectively.  Note that, for any $n$, equation (\ref{eq:variational}) has 36 solutions, which is double the size of the $3 \otimes 3 \otimes 2$ dimensional Hilbert space occupied by $\left| \psi_f(b)\right>$.  In this sense, the basis is overcomplete; for many solutions $\left| \psi_f(b)\right>$, the $(I \otimes I^{\otimes 2j} \otimes \hat{g}_f)(I \otimes I^{\otimes 2(j-1)} \otimes \hat{g}_0)$ part of (\ref{eq:Psifn}) can be rewritten in terms of other solutions $\left| \psi_f^\prime (b)\right>$ and $\left| \psi_f^{\prime\prime}(b)\right>$ in a form such as $(I \otimes I^{\otimes 2j} \otimes \hat{g}_f^\prime)(I \otimes I^{\otimes 2(j-1)} \otimes \hat{g}_f^{\prime\prime})$.

To find the lowest excited energies of the system, we take $2 \pi n/ \ell = 0$ in (\ref{eq:Psifn}).  Solving (\ref{eq:variational}) for $2 \pi n/ \ell = 0$, we find a gap between the ground state and a triplet of excited states with energy $1.3 \times 10^{-11} \epsilon$ at $\theta = 1.56$.  The form of these triplet states is discussed in the Appendix.  Eliminating finite-size effects has reduced the energy some 4 orders of magnitude below the energies appearing Fig.~\ref{fig:gapversuschainlength}.   However, the excited state energy is still non-zero, and our variational framework will yield compelling evidence that the system is gapped.

To see this, we double the size of the ``unit cell'' in (\ref{eq:Psifn}), writing
\begin{eqnarray}
\lefteqn{\left|\Psi^{(2)}_{f,n}\right> = \sqrt{\frac{2}{\ell}} \sum_{j=0}^{(\ell/2)-1} \cos \frac{4\pi n j}{\ell}} \nonumber \\
& & (I \otimes I^{\otimes 2 \ell -2} \otimes \hat{g}_0)(I \otimes I^{\otimes 2 \ell -4} \otimes \hat{g}_0) \dots \nonumber \\
& & (I \otimes I^{\otimes 4j+4} \otimes \hat{g}_0)(I \otimes I^{\otimes 4j+2} \otimes \hat{g}_0) \nonumber \\
& & \hspace{0.5in} (I \otimes I^{\otimes 4j} \otimes \hat{g}^{(2)}_f)  \nonumber \\
& & (I \otimes I^{\otimes 4j-2} \otimes \hat{g}_0)(I \otimes I^{\otimes 4j-4} \otimes \hat{g}_0) \nonumber \\
& & \dots (I \otimes I^{\otimes 2} \otimes \hat{g}_0) (I \otimes \hat{g}_0) \frac{\left|0\right>\left|0\right>+\left|1\right>\left|1\right>}{\sqrt{2}}. \label{eq:Psi2f}
\end{eqnarray}
where we still have $\hat{g}^{(2)}_f = \left| \psi^{(2)}_f(0)\right>\left<0\right|+\left| \psi^{(2)}_f(1)\right>\left<1\right|$, but now $\left| \psi^{(2)}_f(b)\right>$ is defined on a $3 \otimes 3 \otimes 3\otimes 3 \otimes 2$ dimensional space.  We can recover (\ref{eq:Psifn}) by taking 
\begin{eqnarray*}
\lefteqn{ \cos \frac{4\pi n j}{\ell} \hat{g}^{(2)}_f =}\\
& &  \frac{\cos \frac{2\pi n (2 j+1)}{\ell} (I^{\otimes 2} \otimes \hat{g}_f)  \hat{g}_0 + \cos \frac{4\pi n j}{\ell} (I^{\otimes 2} \otimes \hat{g}_0)  \hat{g}_f }{\sqrt{2}}
\end{eqnarray*}
but the form (\ref{eq:Psi2f}) is more general than (\ref{eq:Psifn}) because it allows arbitrary behavior within a doubled unit cell with a $3 \otimes 3 \otimes 3 \otimes 3 \otimes 2$ dimensional Hilbert space rather than within a single unit cell with a $3\otimes 3 \otimes 2$ dimensional Hilbert space.  We minimize the energy to derive an equation analogous to (\ref{eq:variational}) for $\left| \psi^{(2)}_f(b)\right>$.  Solving this analogous equation for $n=0$ at $\theta = 1.56$, we obtain a lowest excited state energy of $5.1 \times 10^{-12} \epsilon$.   

\begin{figure}[htb]
\begin{center}
\includegraphics[width=3.0in]{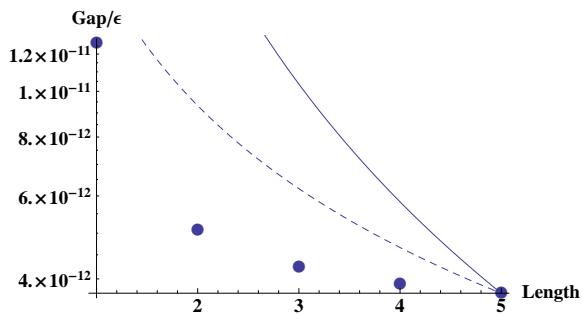}
\end{center}
\caption{Gap to lowest excited state energy as a function of unit cell size obtained by solving (\ref{eq:variational}) for $\left|\Psi_{f,n}\right>$ and analogous equations for $\left|\Psi^{(2)}_{f,n}\right>,\dots,\left|\Psi^{(5)}_{f,n}\right>$.  Lines show that 1/unit cell length (dashed curve) and 1/unit cell length$^2$ (solid curve) behavior cannot account for the data.}
\label{fig:gapversusunitcelllength}
\end{figure}

We then repeat the calculation with a triple-size unit cell, a quadruple-size unit cell, and a quintuple-size unit cell.  The corresponding variational states, $\left|\Psi^{(3)}_{f,n}\right>$, $\left|\Psi^{(4)}_{f,n}\right>$, and $\left|\Psi^{(5)}_{f,n}\right>$ respectively, are less and less constrained, and the corresponding numerical calculations are more and more demanding.  The energies of the lowest excited states are plotted in  
Fig.~\ref{fig:gapversusunitcelllength}.  The data in Fig.~\ref{fig:gapversusunitcelllength} has a weak dependence on unit cell length that provides compelling evidence the gap has a non-zero value even for an infinite chain.  It is instructive to compare with Fig.~\ref{fig:gapversuschainlength} by noting the curves drawn on the two figures.  In Fig.~\ref{fig:gapversuschainlength} for finite chains, the data is fit well by a $1/\ell^2$ curve but in Fig.~\ref{fig:gapversusunitcelllength} the dependence of gap on unit cell length is much weaker.  We conclude that the $1/\ell^2$ dependence in Fig.~\ref{fig:gapversuschainlength} is a finite-size effect that does not reflect the true behavior of an infinite length chain.

\begin{figure}[htb]
\begin{center}
\includegraphics[width=3.0in]{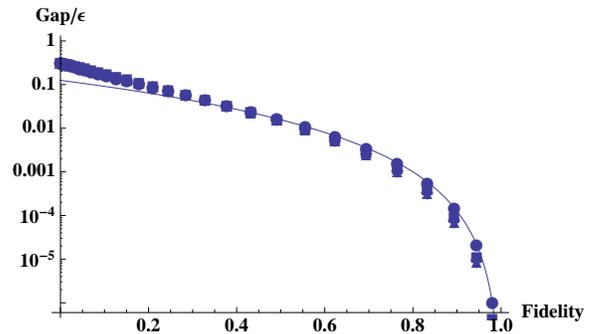}
\end{center}
\caption{Gap to lowest excited state energy as a function of fidelity $f(\theta)$ defined in (\ref{eq:f}).  Points are obtained by performing 5 variational calculations for $\left|\Psi_{f,n=0}\right>,\dots,\left|\Psi^{(5)}_{f,n=0}\right>$ at each $\theta$.  The results for $\left|\Psi_{f,n=0}\right>\dots,\left|\Psi^{(5)}_{f,n=0}\right>$ are designated by $\bullet$,  $\blacksquare$,  $\blacklozenge$, $\blacktriangle$, and $\blacktriangledown$ respectively (the 5 symbols tend to overlap one another on the logarithmic scale of the figure).  The solid curve, given by $(1/8)(1-f(\theta))^3$, fits the points well.}
\label{fig:gapversusfidelity}
\end{figure}

Based on Lieb-Robinson arguments \cite{Hastings06}, the fidelity of a Bell pair formed at ends of a chain must be small when the energy gap of the chain is large.  This is evident in the calculations presented in Fig.~\ref{fig:gapversusfidelity}.  

\section{Relationship to AKLT and quantum buses}

It is instructive to compare our parent spin Hamiltonian and ground state to those of the AKLT model \cite{Affleck1987,Affleck1988,Affleck1989}.  The AKLT ground state is a ``valence-bond solid" that can be obtained by imagining a chain of sites with 2 spin-1/2 particles on each site.  At site $i$ of the chain, we arbitrarily choose one of the spin-1/2 particles and form a singlet between it and a spin-1/2 particle at site $i+1$; we then form a singlet between the remaining spin-1/2 particle at site $i$ and a spin-1/2 particle at site $i-1$.  We symmetrize the resulting state over all such arbitrary choices at every site $i$.  Equivalently, instead of symmetrizing, we project out the singlet part of the state at each site $i$, leaving only the symmetric spin triplet part.  The result is a state with an effective spin-1 particle $S_i$ at each site $i$.  The parent spin Hamiltonian is obtained by projecting adjacent sites $i$ and $i+1$ on to spin 2
\begin{eqnarray}
\lefteqn{H_{AKLT}  =  \sum_i P^{(2)}(S_i+S_{i+1})} \nonumber \\
& \propto & \sum_i S_i \cdot S_{i+1} + \frac{1}{3}(S_i \cdot S_{i+1})^2 + \text{const}.
\end{eqnarray}
This Hamiltonian annihilates the AKLT ground state: of the 4 spin-1/2 particles at sites $i$ and $i+1$, 2 of them are bound into a spin singlet with total spin 0 and the remaining 2 spin-1/2 particles can produce a total spin of at most 1.  Thus, the projection on to spin 2 vanishes.  AKLT have proven that this Hamiltonian is gapped \cite{Affleck1987,Affleck1988}.

Suppose that we change our convention in eq. (\ref{eq:psi(b)}) so that  $(\left|0\right>\otimes \left|0\right>+\left|1\right>\otimes \left|1\right>)/ \sqrt{2}$ is replaced with the singlet state $(\left|0\right>\otimes \left|1\right>-\left|1\right>\otimes \left|0\right>)/ \sqrt{2}$.  We change the Hamiltonians (\ref{eq:HCreatepair}) and (\ref{eq:HProjection}) accordingly.  Examining eq. (\ref{eq:HProjection}), we then see that it transitions the singlet part of 2 spin-1/2 particles on to the state $\left|{\scriptscriptstyle IDLE}\right>\left|{\scriptscriptstyle IDLE}\right>$.  This is reminiscent of the step during the construction of the AKLT ground state in which we project out the singlet part of the 2 spin-1/2 particles at each site $i$.  However, the effect of including eq. (\ref{eq:HProjection}) in our Hamiltonian is actually to enhance the contribution of the singlet state rather than to project it out.  As a result, our parent spin Hamiltonian emulates correct teleportation in the limit near $\theta = \pi/2$, and the fidelity (\ref{eq:f}) falls off relatively slowly.  In contrast, the AKLT ground state emulates failed teleportation in which projecting out the singlet state corresponds to Bell measurement of a triplet state, but no Pauli operator gets applied to correct the teleported state.  As a result, AKLT correlations fall off very rapidly \cite{Affleck1989}.

It is worth noting that, within the $3 \otimes 3$ Hilbert spaces along the spin chain in Fig.~\ref{fig:spinchain}, $4$ of $9$ states have the form $\left|{\scriptscriptstyle IDLE}\right>\left|0\right>$, $\left|{\scriptscriptstyle IDLE}\right>\left|1\right>$, $\left|0\right>\left|{\scriptscriptstyle IDLE}\right>$, $\left|1\right>\left|{\scriptscriptstyle IDLE}\right>$ and are energetically penalized by the sums in (\ref{eq:HProjection}).  The 5 remaining states $\left|{\scriptscriptstyle IDLE}\right>\left|{\scriptscriptstyle IDLE}\right>$, $\left|0\right>\left|0\right>$, $\left|0\right>\left|1\right>$, $\left|1\right>\left|0\right>$, $\left|1\right>\left|1\right>$ can be regarded as forming an effective spin-2 particle at each``site" of our spin chain.  The fact that our parent spin Hamiltonian is gapped then seems consistent with Haldane's conjecture \cite{Haldane1983}.  

There has been some interest in using spin chains as quantum information buses \cite{Bose2003,Christandl2004,Verstraete2004,Wojcik2005,Venuti2006,Venuti2007,Cappellaro2007,Ferreira2008,Friesen2007,Wang2009,Wu2009,Oh2010,Yao2010,Banchi2011,Oh2011,Yang2011,Pemberton2011,Shim2011,Oh2012}, and it is worth considering the relationship of our parent spin Hamiltonian to this work.  In particular, we note that the AKLT Hamiltonian \cite{Verstraete2004} has appeared in a proposal for a quantum channel in which measurement of the spin-1 particles in the AKLT chain and application of Pauli operation corrections allows teleportation of a quantum state along the chain.  This is closely related to our statement above that the AKLT ground state enacts an time-independent emulation of failed teleportation.

One can imagine employing our Hamiltonian as a sort of quantum bus to passively produce Bell pairs.  For this purpose, it would make sense to choose the value of $\theta$ to be as close as possible to $\pi/2$ while keeping the gap large enough to stave off thermal excitations.  Since the fidelity  of the chain's Bell pair is $f(\theta)^\ell \sim (1 - \ell (1-f(\theta)))$, the maximum practical length $\ell$ of the chain would then be constrained to around $1/(1-f(\theta))$.  This bus would function quite differently than the channel of \cite{Verstraete2004} since it would have limited fidelity but would require no active measurement to serve up a Bell pair separated by the chain length.

\section{Conclusion}

We have proposed a spin chain with a ground state that emulates teleportation and a parent spin Hamiltonian.  Guided by exact diagonalization results, we framed a mean field theory to argue that the parent spin Hamiltonian is gapped.  We pointed out a revealing connection to the AKLT model \cite{Affleck1987,Affleck1988,Affleck1989} and to a quantum channel proposal related to the AKLT model \cite{Verstraete2004}.  We also noted that including a transition to an extra state in the parent Hamiltonian, as in (\ref{eq:HProjection}), provides a means of amplifying part of the ground state.  This is an interesting technique that complements the spin projection used to project out part of the AKLT ground state.  Various generalizations of the AKLT model exist, including higher dimensional spin lattices \cite{Affleck1987}, and one expects that generalizations of our spin chain are similarly possible.

\section*{Acknowledgements}

The author thanks Charles Tahan and Yun-Pil Shim for comments.

\bibliography{Gapped}

\section{Appendix}

In this brief appendix, we convey intuition about the form of the lowest excited states of the mean field calculation (\ref{eq:variational}).  For simplicity, we focus on the single unit cell calculation rather than on the calculations with larger unit cells.  For $\theta \rightarrow \pi/2$, the triplet of excited states has the form 
\begin{eqnarray*}
\lefteqn{\left| \psi_1(0)\right>  =}  \nonumber \\
&&\left|0\right> \otimes \frac{\left|0\right> \otimes \left|1\right> +  \left|1\right> \otimes \left|0\right>}{2} - \left|1\right> \otimes \frac{\left|0\right> \otimes \left|0\right>  +  \left|1\right> \otimes \left|1\right>}{2}, \\
\lefteqn{\left| \psi_1(1)\right> =} \nonumber \\
&& \left|1\right> \otimes \frac{\left|0\right> \otimes \left|1\right> + \left|1\right> \otimes \left|0\right>}{2}  - \left|0\right> \otimes \frac{\left|0\right> \otimes \left|0\right> +   \left|1\right> \otimes \left|1\right>}{2}, \\
\lefteqn{\left| \psi_2(0) \right> =} \nonumber \\
& &  \left|0\right> \otimes \frac{\left|0\right> \otimes \left|1\right> -  \left|1\right> \otimes \left|0\right>}{2} - \left|1\right> \otimes \frac{\left|0\right> \otimes \left|0\right>  +  \left|1\right> \otimes \left|1\right>}{2}, \\
\lefteqn{\left| \psi_2(1) \right>  = } \nonumber \\
& & \left|1\right> \otimes \frac{\left|0\right> \otimes \left|1\right> - \left|1\right> \otimes \left|0\right>}{2}  + \left|0\right> \otimes \frac{\left|0\right> \otimes \left|0\right> +   \left|1\right> \otimes \left|1\right>}{2},\\
\lefteqn{\left| \psi_3(0) \right> = } \nonumber \\
& & \left|0\right> \otimes \frac{\left|0\right> \otimes \left|0\right> - \left|1\right> \otimes \left|1\right>}{2}  - \left|0\right> \otimes \frac{\left|0\right> \otimes \left|0\right>  +  \left|1\right> \otimes \left|1\right>}{2}, \\
\lefteqn{\left| \psi_3(1) \right> = } \nonumber \\
& & \left|1\right> \otimes \frac{\left|0\right> \otimes \left|0\right> - \left|1\right> \otimes \left|1\right>}{2}  + \left|1\right> \otimes \frac{\left|0\right> \otimes \left|0\right> +   \left|1\right> \otimes \left|1\right>}{2},
\end{eqnarray*}
corresponding to a bit-flip within the Bell pair such that $\left| \psi_1(b)\right> = (I\otimes I \otimes X - X \otimes I \otimes I) \left|b\right> \otimes \left(\left|0\right> \otimes \left|0\right> +  \left|1\right> \otimes \left|1\right>\right)/2$, a phase flip such that $\left| \psi_3(b)\right> = (I\otimes I \otimes Z - Z \otimes I \otimes I) \left|b\right> \otimes \left(\left|0\right> \otimes \left|0\right> +  \left|1\right> \otimes \left|1\right>\right)/2$, or a combined bit-and-phase flip such that $\left| \psi_2(b)\right> = (I\otimes I \otimes XZ - XZ \otimes I \otimes I) \left|b\right> \otimes \left(\left|0\right> \otimes \left|0\right> +  \left|1\right> \otimes \left|1\right>\right)/2$.

\end{document}